\documentclass[trackchanges]{aastex701}

\usepackage{makecell}
\usepackage{graphicx}
\usepackage{subcaption}
\usepackage{caption} 
\usepackage{booktabs}
\usepackage{amssymb}
\usepackage{amsmath}
\usepackage{float}
\usepackage{xtab}
\usepackage{orcidlink}
\usepackage{soul}

\begin{document}

\title{Dynamical Dark Energy and the Unresolved Hubble Tension: Multi-model Constraints\\ from DESI 2025 and Other Probes}

\correspondingauthor{Yun Chen}
\email{chenyun@bao.ac.cn}

\author[orcid=0000-0002-4518-6035]{Zhuoming Zhang}
\affiliation{National Astronomical Observatories, Chinese Academy of Sciences\\
Beijing 100101, China}
\affiliation{College of Astronomy and Space Sciences, University of Chinese Academy of Sciences\\
Beijing, 100049, China}
\email[show]{}

\author[orcid=0000-0002-9855-2342]{Tengpeng Xu}
\affiliation{School of Physics and Astronomy, Sun Yat-Sen University,\\ Zhuhai 519082, China}
\email[show]{}

\author[orcid=0000-0001-8919-7409]{Yun Chen}
\affiliation{National Astronomical Observatories, Chinese Academy of Sciences\\
Beijing 100101, China}
\affiliation{College of Astronomy and Space Sciences, University of Chinese Academy of Sciences\\
Beijing, 100049, China}
\email[show]{}

\begin{abstract}
We present a Bayesian comparative analysis of five cosmological models: $\Lambda$CDM, $w$CDM, $w_0w_a$CDM, $\phi$CDM (with scalar-field dark energy), and an interacting dark energy scenario (the $\xi$-index model), to investigate dark energy evolution and the Hubble tension. Utilizing the latest data from the Dark Energy Spectroscopic Instrument (DESI) DR2 (Baryon Acoustic Oscillations, BAO), Pantheon+ (Type Ia Supernovae, SNIa), and Cosmic Microwave Background (CMB) data (including lensing) from \textit{Planck} and the Atacama Cosmology Telescope (ACT), we report three key findings. First, the Hubble constant ($H_0$) inferred from the combined data consistently aligns with early-universe measurements across all models, indicating a persistent Hubble tension. Second, we find compelling evidence for dynamical dark energy: early-universe (CMB) constraints favor a phantom phase (with an equation-of-state parameter $w < -1$), while late-universe (BAO/SNIa) data prefer quintessence ($w > -1$). Third, the full dataset suggests a late-time interaction between dark energy and matter. Our results demonstrate that dark energy evolves with cosmic time, challenging the cosmological constant paradigm.
\end{abstract}


\section{Introduction} \label{sec:intro}
The discovery of the cosmic expansion in the early 20th century established the expansion history as a fundamental probe of cosmic composition and physical properties \citep{1929PNAS...15..168H}. Subsequent observations of Type Ia supernovae (SNIa) not only confirmed this expansion but also revealed its accelerating nature \citep{riess1998observational, perlmutter1999measurements}. This acceleration has been robustly confirmed by multiple independent probes \citep{weinberg2013observational}, including cosmic microwave background (CMB) anisotropies \citep{spergel2003first,2011ApJS..192...18K, 2020A&A...641A...6P,madhavacheril2024atacama}, baryon acoustic oscillations (BAO) \citep{percival2007measuring,2011MNRAS.416.3017B,2012PhRvD..86j3518P,2017MNRAS.470.2617A,alam2021completed}, and weak gravitational lensing \citep{bartelmann2001weak,2015MNRAS.450.2888L,2018MNRAS.474.1116S,2018ARA&A..56..393M,2021A&A...646A.140H,2022PhRvD.105b3520A}.

Within the standard $\Lambda$CDM cosmological model, this phenomenon is attributed to a cosmological constant—a static form of dark energy with a constant equation-of-state (EoS) parameter $w = -1$ \citep{1984ApJ...284..439P, 2012IJMPD..2130002Y}. Despite its remarkable success in describing a wide range of cosmological observations, the physical origin of dark energy remains one of the most profound puzzles in modern physics \citep{1989RvMP...61....1W}. This has motivated a rich spectrum of theoretical extensions beyond $\Lambda$CDM model \citep{2003RvMP...75..559P, 2006IJMPD..15.1753C, 2008ARA&A..46..385F, 2009ARNPS..59..397C, 2010PhLB..690..333X, 2011PhLB..698..175C, 2013FrPhy...8..828L, 2020PhR...857....1F, 2025arXiv250200923V}.

Recent observations from the Dark Energy Spectroscopic Instrument (DESI) have provided intriguing evidence suggesting that dark energy may not be a simple cosmological constant. \citet{2025PhRvD.112h3515A} reported constraints on the EoS within the $w_0w_a$CDM framework that favor $w_0 > -1$ and $w_a < 0$, indicating potential dynamical behavior. Concurrently, \citet{gu2025dynamical} employed non-parametric methods to reveal significant redshift evolution in the EoS of dark energy. These findings directly challenge the foundational assumption of the $\Lambda$CDM model and call for a systematic investigation of alternative dark energy scenarios.

A persistent challenge to the standard cosmological model is the Hubble tension—a statistically significant discrepancy between early- and late-universe measurements of the Hubble constant $H_0$. The CMB-based estimate from \citet{collaboration2020planck} ($67.40 \pm 0.50$ km/s/Mpc) exhibits clear tension with the Cepheid-calibrated distance-ladder result from \citet{riess2022comprehensive} ($73.04 \pm 1.04$ km/s/Mpc). The emergence of potential dark energy dynamics, coupled with the model-dependent nature of $H_0$ inference, makes it imperative to test whether the Hubble tension can be alleviated within a broader class of cosmological models, or if it points to more fundamental systematics or new physics.

In this work, we conduct a systematic comparison of five representative cosmological frameworks using the latest observational data, including DESI DR2 BAO, the Pantheon+ SNIa sample, and \textit{Planck} 2018 and Atacama Cosmology Telescope (ACT) DR6 CMB and lensing data. The investigated models are strategically chosen to span key theoretical directions: the standard $\Lambda$CDM paradigm, the $w$CDM model (constant EoS), the $w_0w_a$CDM model (dynamical EoS), the $\phi$CDM model (scalar-field dark energy), and an interacting dark energy scenario ($\xi$-index model).

This study is grounded in a framework of multiple complementary perspectives. First, we perform Bayesian model comparison to quantify the relative empirical support for each model. Second, we derive constraints on $H_0$ under each framework to evaluate their potential for resolving the Hubble tension. Furthermore, we analyze the redshift evolution of the dark energy EoS to robustly test for dynamics. Finally, we place constraints on the coupling in the $\xi$-index model to probe interactions between the dark energy and matter.

This paper is organized as follows. Section \ref{sec:models} introduces the theoretical frameworks of the five cosmological models. Section \ref{sec:datasets} describes the observational datasets. Section \ref{sec:results} presents our methodology, results, and discussions. Section \ref{sec:conclusion} summarizes our principal findings and conclusions.

\section{Theoretical Models} \label{sec:models}

\subsection{The $\Lambda$CDM Model and Its Minimal Extensions}
Within the framework of general relativity, the cosmic expansion history is described by the Friedmann equations, with dark energy playing a crucial role in the late-time acceleration. For a non-interacting dark energy component characterized by an EoS parameter $w(z)$, the evolution of its energy density follows the continuity equation, yielding the dimensionless density evolution:

\begin{equation}
\frac{\Omega_{\mathrm{DE}}(z)}{\Omega_{\mathrm{DE}, 0}} = \exp \left[3 \int_0^z \left[1 + w(z^{\prime})\right] \frac{d z^{\prime}}{1 + z^{\prime}}\right],
\label{eq:Omega_de}
\end{equation}

where $\Omega_{\mathrm{DE},0}$ denotes the present-day dark energy density parameter. This fundamental relation forms the basis for a hierarchy of phenomenological dark energy parameterizations, which we systematically examine below.

\begin{itemize}
\item \textbf{The $\Lambda$CDM Model}

As the standard cosmological paradigm, the $\Lambda$CDM model attributes cosmic acceleration to a cosmological constant with constant energy density \citep{1984ApJ...284..439P}. This corresponds to a fixed EoS parameter $w(z) = -1$ across all redshifts, resulting in:
\begin{equation}
\Omega_{\mathrm{DE}}(z) = \Omega_{\mathrm{DE}, 0}.
\end{equation}
The model's remarkable success in fitting diverse observational datasets establishes it as the fundamental reference for testing extensions.

\item \textbf{The $w$CDM Model}

This minimal extension introduces a constant but non-unity EoS parameter $w = \mathrm{Constant}$ \citep{1991PhRvD..43.3802R}, allowing quantitative assessment of deviations from the cosmological constant scenario. The dark energy density evolves as:
\begin{equation}
\Omega_{\mathrm{DE}}(z) = \Omega_{\mathrm{DE}, 0}(1+z)^{3(1+w)}.
\end{equation}
This parametrization provides a straightforward framework for constraining persistent deviations from $w = -1$.

\item \textbf{The $w_0w_a$CDM Model}

To capture potential dynamical evolution, the Chevallier-Polarski-Linder (CPL) parametrization \citep{chevallier2001accelerating, linder2003exploring} introduces redshift dependence through $w(z) = w_0 + w_a\frac{z}{1+z}$, where $w_0$ represents the present value and $w_a$ characterizes temporal variation. The corresponding density evolution is given by:
\begin{equation}
\Omega_{\mathrm{DE}}(z) = \Omega_{\mathrm{DE}, 0} (1+z)^{3(1 + w_0 + w_a)} \exp\left(-3 w_a \frac{z}{1+z}\right).
\end{equation}
This flexible two-parameter formulation enables comprehensive investigation of dark energy dynamics across cosmic history.
\end{itemize}

\subsection{Scalar Field Dark Energy: The $\phi$CDM Framework}
The $\phi$CDM model provides a physically motivated framework for dynamical dark energy by introducing a canonical scalar field $\phi$ evolving in a self-interaction potential $V(\phi)$. Unlike purely phenomenological parameterizations, this approach offers a microphysical origin for cosmic acceleration, with the field's dynamics naturally generating a time-varying EoS.

The fundamental action in the context of general relativity is given by:
\begin{equation}
S = \int d^4x \sqrt{-g} \left( -\frac{m_{\mathrm{P}}^2}{16\pi} R + \mathcal{L}{\mathrm{m,r}} + \mathcal{L}{\phi} \right),
\end{equation}
where $g_{\mu\nu}$ denotes the metric tensor, $R$ is the Ricci scalar, $m_{\mathrm{P}}$ represents the Planck mass, $\mathcal{L}{\mathrm{m,r}}$ encompasses the matter and radiation sectors, and the scalar field Lagrangian takes the form:
\begin{equation}
\mathcal{L}{\phi} = \frac{1}{2} g^{\mu\nu} \partial_{\mu}\phi \partial_{\nu}\phi - V(\phi).
\end{equation}

In a spatially flat Friedmann-Lema\^itre-Robertson-Walker (FLRW) universe, the energy density and pressure associated with the scalar field are expressed as:
\begin{eqnarray}
\rho_{\phi} &=& \frac{1}{2} \dot{\phi}^2 + V(\phi), \\
p_{\phi} &=& \frac{1}{2} \dot{\phi}^2 - V(\phi),
\end{eqnarray}
which yields a dynamically evolving EoS parameter:
\begin{equation}
w_{\phi} = \frac{p_{\phi}}{\rho_{\phi}} = \frac{\dot{\phi}^2 - 2V(\phi)}{\dot{\phi}^2 + 2V(\phi)}.
\end{equation}
This parameter naturally remains bounded within the range $-1 \leq w_{\phi} \leq 1$, where the lower limit corresponds to a potential-dominated state ($\dot{\phi}^2 \ll V(\phi)$) and the upper limit to a kinetic-dominated state ($\dot{\phi}^2 \gg V(\phi)$).

The temporal evolution of the field is governed by the Klein-Gordon equation in an expanding universe:
\begin{equation}
\ddot{\phi} + 3H\dot{\phi} + \frac{dV}{d\phi} = 0,
\end{equation}
while the background expansion history follows from the Friedmann equation:
\begin{equation}
H^2 = \frac{8\pi}{3m_{\mathrm{P}}^2} \left( \rho_{\mathrm{m}} + \rho_{\phi} \right),
\end{equation}
with the matter component evolving as $\rho_{\mathrm{m}} = \rho_{\mathrm{m0}} a^{-3}$.

We adopt the well-motivated inverse power-law potential \citep{peebles1988cosmology, ratra1988cosmological}:
\begin{equation}
V(\phi) = \hat{V} \phi^{-\alpha},
\end{equation}
where $\alpha \geq 0$ parameterizes the steepness of the potential and $\hat{V}$ is a dimensional constant. This minimal parameterization encompasses the cosmological constant as the special case $\alpha = 0$, while $\alpha > 0$ generates a rich spectrum of dynamical dark energy behavior. The complete cosmological evolution is determined through numerical integration of the coupled Friedmann and Klein-Gordon equations, providing a self-consistent framework for confronting scalar field dynamics with observational data.

\subsection{Interacting Dark Energy: The $\xi$-index Model}
\label{ch:xi-index model}
The $\xi$-index model provides a phenomenological framework for investigating potential interactions between dark energy and matter \citep{dalal2001testing, chen2010using}. The model is built upon a fundamental scaling relation between their energy densities:
\begin{equation}
\rho_\mathrm{X} = \kappa\rho_\mathrm{m} a^{\xi}, \quad \text{or equivalently} \quad \Omega_\mathrm{X}(a) = \kappa \Omega_\mathrm{m}(a) a^{\xi},
\end{equation}
where $\rho_\mathrm{X}$ and $\rho_\mathrm{m}$ denote the energy densities of dark energy and matter, $\Omega_\mathrm{X}$ and $\Omega_\mathrm{m}$ their corresponding dimensionless density parameters, and $\kappa \equiv \rho_\mathrm{X,0}/\rho_\mathrm{m,0}$ is the present-day density ratio. This parameterization generalizes two well-known limits: it reduces to $\rho_\mathrm{X} \propto \rho_\mathrm{m} a^{3}$ for the $\Lambda$CDM limit ($\xi=3$), and to $\rho_\mathrm{X} \propto \rho_\mathrm{m}$ for a self-similar scaling that mitigates the coincidence problem ($\xi=0$). The exponent $\xi$ thus quantifies deviations from the non-interacting scenario, motivating the name of the model.

The total energy density of the coupled sector is $\rho_{\mathrm{mix}} = \rho_\mathrm{X} + \rho_\mathrm{m}$, allowing the dark energy density to be expressed as
\begin{equation}
\rho_\mathrm{X} = \frac{\rho_{\mathrm{mix}} \kappa a^{\xi}}{1+\kappa a^{\xi}}.
\label{eq:Omega_X}
\end{equation}

The conservation equation for dark energy and matter follows from the continuity equation in the FLRW metric:
\begin{equation}
\frac{\mathrm{d} \rho_{\mathrm{mix}}}{\mathrm{d} a} + \frac{3}{a}\left(\rho_{\mathrm{mix}} + w_\mathrm{X} \rho_\mathrm{X}\right) = 0,
\label{eq:energy_conservation}
\end{equation}
where $w_\mathrm{X}$ is the dark energy EoS parameter. This conservation law can be decomposed into individual continuity equations for each component, revealing an explicit interaction term $Q$:
\begin{equation}
\frac{\mathrm{d} \rho_\mathrm{m}}{\mathrm{d} a} + \frac{3}{a} \rho_\mathrm{m} = Q, \quad \frac{\mathrm{d} \rho_\mathrm{X}}{\mathrm{d}a} + \frac{3}{a}(1+w_\mathrm{X}) \rho_\mathrm{X} = -Q,
\end{equation}
where the interaction term is given by
\begin{equation}
Q = -\frac{(\xi + 3w_\mathrm{X})\rho_\mathrm{m} \kappa a^{\xi-1}}{1+\kappa a^{\xi}}.
\end{equation}

The sign of the quantity $(\xi + 3w_\mathrm{X})$ determines the direction of energy transfer:
\begin{itemize}
\item $\xi + 3w_\mathrm{X} > 0$: Energy flows from matter to dark energy ($Q < 0$).
\item $\xi + 3w_\mathrm{X} < 0$: Energy transfers from dark energy to matter ($Q > 0$).
\item $\xi + 3w_\mathrm{X} = 0$: No interaction occurs ($Q = 0$), recovering the non-interacting cosmology.
\end{itemize}

Integration of the conservation equation (\ref{eq:energy_conservation}) yields the evolution of the total density:
\begin{equation}
\frac{\rho_{\mathrm{mix}}(a)}{\rho_{\mathrm{mix},0}} = \exp\left[-3\int_{1}^{a} \frac{\mathrm{d}a'}{a'}\left(1 + \frac{w_\mathrm{X} \kappa a'^{\xi}}{1+\kappa a'^{\xi}}\right)\right].
\label{eq:rho_ratio}
\end{equation}

For constant parameters $w_\mathrm{X}$ and $\xi$, the integral can be solved analytically. The resulting dimensionless Hubble parameter in a flat universe is:
\begin{equation}
E^{2}(a) = \frac{H^{2}(a)}{H_0^{2}} = \Omega_{\mathrm{mix},0}a^{-3}\left(\frac{1+\kappa}{1+\kappa a^{\xi}}\right)^{\frac{3w_\mathrm{X}}{\xi}} + \Omega_{\mathrm{r},0}a^{-4},
\label{eq:Hubble_param}
\end{equation}
where $\Omega_{\mathrm{mix},0}$ and $\Omega_{\mathrm{r},0}$ are the present-day density parameters for the coupled sector and radiation, respectively. The model is thus characterized by the parameter set $\mathbf{p} = \left(\Omega_{\mathrm{mix},0}, \Omega_\mathrm{r,0}, \kappa, w_\mathrm{X}, \xi\right)$.

This formulation provides a self-consistent framework for constraining dark energy--matter interactions through cosmological observations. The $\xi$ parameter directly probes deviations from the non-interacting scenario, offering valuable insights into the nature of the dark sector and the cosmic coincidence problem.

\section{Observational Data} \label{sec:datasets}

\subsection{Type Ia Supernovae (Pantheon+ Sample)}
\label{sec:snia}

SNIa serve as fundamental standardizable candles for probing the cosmic expansion history, particularly at low redshifts where BAO measurements are limited by cosmic variance. We utilize the Pantheon+ sample \citep{scolnic2022Pantheon+, brout2022Pantheon+}, which comprises 1,550 spectroscopically confirmed SNIa spanning the redshift range $z = 0.001$--2.26. Following \citet{brout2022Pantheon+}, we use the $z > 0.01$ subset for cosmological analysis to avoid potential systematics from uncertain peculiar velocity corrections.

The Pantheon+ collaboration employs the ``Bayesian Estimation Applied to Multiple Species (BEAMS) with Bias Corrections (BBC)'' method \citep{kessler2017correcting} to calibrate the SNIa distances and correct for various biases. The key observable is the corrected apparent magnitude, defined as:
\begin{equation}
m_{\mathrm{obs}} \equiv m_{\mathrm{B}} + \alpha x_1 - \beta c - \delta_{\mathrm{bias}} + \delta_{\mathrm{host}},
\label{eq:SNIa}
\end{equation}
where $m_{\mathrm{B}}$ is the observed B-band apparent magnitude. The parameters $\alpha$ and $\beta$ are global nuisance parameters characterizing the stretch-luminosity and color-luminosity relationships, with $x_1$ and $c$ quantifying the light-curve stretch and color, respectively. The terms $\delta_{\mathrm{bias}}$ and $\delta_{\mathrm{host}}$ correct for selection biases \citep{popovic2021improved} and host-galaxy mass correlations.

The theoretical distance modulus is computed as:
\begin{equation}
\mu_{\mathrm{th}}(\mathbf{p}; z) = 5 \log_{10}\left[\frac{D_{\mathrm{L}}(\mathbf{p}; z)}{\mathrm{Mpc}}\right] + 25,
\end{equation}
where the luminosity distance $D_{\mathrm{L}}$ in a flat universe is related to the line-of-sight comoving distance $D_{\mathrm{M}}$ by:
\begin{equation}
D_{\mathrm{L}}(\mathbf{p}; z) = (1+z) D_{\mathrm{M}}(\mathbf{p}; z) = c_{0}(1+z) \int_0^z \frac{dz'}{H(\mathbf{p}; z')},
\end{equation}
where $c_0$ denotes the speed of light. The final cosmological constraints are derived from the likelihood $\mathcal{L}_{\mathrm{SNIa}} \propto \exp(-\chi^2_{\mathrm{SNIa}}/2)$, where:
\begin{equation}
\chi^2_{\mathrm{SNIa}} = \Delta\vec{m}^T C_{\mathrm{stat+syst}}^{-1} \Delta\vec{m}.
\end{equation}
Here, $\Delta m_i = m_{\mathrm{obs},i} - \mu_{\mathrm{th}}(\mathbf{p}; z_i)$ is the residual vector, and $C_{\mathrm{stat+syst}}$ is the full covariance matrix incorporating both statistical and systematic uncertainties. The Pantheon+ data are publicly available\footnote{\url{https://github.com/PantheonPlusSH0ES/DataRelease}}.

\subsection{Baryon Acoustic Oscillations (DESI DR2)}
We utilize BAO measurements from DESI DR2 \citep{2025PhRvD.112h3515A, 2025PhRvD.112h3514A}, covering the redshift range $0.295 \le z_{\mathrm{eff}} \le 2.330$. DESI DR2 represents a significant improvement over DR1, incorporating three years of observations with doubled Lyman-$\alpha$ forest spectra and substantially enhanced signal-to-noise ratios. This data quality improvement enables the most precise measurement of the BAO scale at $z > 2$ to date, achieving $0.65\%$ statistical precision \citep{2025PhRvD.112h3514A}.

DESI DR2 employs multiple tracer populations: the bright galaxy sample (BGS; $z_{\mathrm{eff}}=0.295$), luminous red galaxies (LRG; $z_{\mathrm{eff}}=0.510, 0.706, 0.922$), emission line galaxies (ELG; $z_{\mathrm{eff}}=0.955, 1.321$), a combined LRG-ELG sample ($z_{\mathrm{eff}}=0.934$), quasi-stellar objects (QSO; $z_{\mathrm{eff}}=1.484$), and the Lyman-$\alpha$ forest (Lya; $z_{\mathrm{eff}}=2.330$). The BAO observables comprise: the comoving angular diameter distance $D_{\mathrm{M}} / r_{\mathrm{d}}$, the Hubble distance $D_{\mathrm{H}} / r_{\mathrm{d}}$, and the volume-averaged distance $D_{\mathrm{V}} / r_{\mathrm{d}}$, where $r_{\mathrm{d}}$ is the sound horizon at the drag epoch. The theoretical expressions are:

\begin{equation}
r_\mathrm{d}(\mathbf{p})=\int_{z_\mathrm{d}}^\infty\frac{c_\mathrm{s}(\mathbf{p}; z)}{H(\mathbf{p}; z)}dz,
\end{equation}
\begin{equation}
D_\mathrm{H}(\mathbf{p}; z)=\frac{c_0}{H(\mathbf{p}; z)},
\end{equation}
\begin{equation}
D_\mathrm{V}(\mathbf{p}; z)\equiv\left[zD_\mathrm{M}(z)^2D_\mathrm{H}(z)\right]^{1/3},
\end{equation}
where $c_\mathrm{s}(\mathbf{p}; z)$ is the sound speed in the photon-baryon fluid, and $z_{\mathrm{d}} \approx 1060$ is the redshift at the drag epoch.

We use data from Table~IV of \citet{2025PhRvD.112h3515A}. While that work only reports $D_\mathrm{V}/r_\mathrm{d}$ for the BGS, we adopt its $D_\mathrm{M}/r_\mathrm{d}$ and $D_\mathrm{H}/r_\mathrm{d}$ values for the other tracers, in line with the study’s own reliance on these two quantities. This consolidated dataset serves two main purposes: first, to allow a direct comparison with the $\Lambda$CDM model and to validate our methodology; second, to facilitate a clear comparison of discrepancies between models and to avoid potential inconsistencies that could arise from using heterogeneous data. The key elements of the dataset are summarized below:
\begin{itemize}
\item $D_\mathrm{V}/r_\mathrm{d}$ at $z_{\mathrm{eff}}=0.295$ (BGS)
\item $(D_\mathrm{M}/r_\mathrm{d}, D_\mathrm{H}/r_\mathrm{d})$ at $z_{\mathrm{eff}}=0.510, 0.706, 0.934, 1.321, 1.484, 2.330$ (LRG, ELG, QSO, Lyman-$\alpha$)
\end{itemize}
This yields a total of 13 BAO data points.

The BAO likelihood follows $\mathcal{L}_{\mathrm{BAO}} \propto \exp(-\chi^2_{\mathrm{BAO}}/2)$, with
\begin{equation}
\chi_{\mathrm{BAO}}^2=\Delta\vec{D}^T C_{\mathrm{BAO}}^{-1}\Delta\vec{D},
\end{equation}
where $\Delta\vec D_i=\vec D_{\mathrm{obs},i}-\vec D_{\mathrm{th}}(\mathbf{p}; z_i)$ and $C_{\mathrm{BAO}}$ is the covariance matrix.

\subsection{Cosmic Microwave Background and Lensing Data}
We utilize precision measurements of CMB temperature and polarization anisotropies from the \textit{Planck} 2018 data release, complemented by CMB lensing potential power spectrum reconstructions from both \textit{Planck} and the ACT DR6. This comprehensive dataset, referred to as \textbf{``CMB+lensing''}, provides robust constraints on cosmological parameters through multiple complementary observables.

The \textit{Planck} likelihood is constructed from three primary components:
\begin{itemize}
    \item \textbf{Low-$\ell$ temperature (TT)}: Covering multipoles $2 \leq \ell \leq 29$, derived from the \textsc{Commander} component-separated maps across the 30--857 GHz frequency range \citep{akrami2020planck}.
    
    \item \textbf{Low-$\ell$ polarization (EE)}: Spanning $2 \leq \ell \leq 29$, obtained from 100-GHz and 143-GHz maps using the \textsc{SimAll} likelihood methodology \citep{2020A&A...641A...6P}, with rigorous validation against end-to-end simulations.
    
    \item \textbf{High-$\ell$ spectra (TT, TE, EE)}: Employing the \textsc{Plik} likelihood \citep{aghanim2020plancka}, with TT covering $30 \leq \ell \leq 2508$ and TE/EE spectra extending to $\ell \leq 1996$. These measurements are derived from cross-correlation analyses of half-mission data at 100-GHz, 143-GHz, and 217-GHz frequency channels.
\end{itemize}

The CMB lensing likelihood incorporates the joint reconstruction from \textit{Planck} NPIPE PR4 and ACT DR6 observations \citep{aghanim2020planckc, qu2024atacama, madhavacheril2024atacama}, providing complementary constraints on the matter distribution through gravitational lensing. This combined dataset significantly enhances the constraining power on cosmological parameters, particularly those governing dark energy evolution and cosmic structure growth.

The complete CMB+lensing likelihood function integrates these independent measurements, enabling precise determination of fundamental cosmological parameters while accounting for systematic uncertainties and covariance between different observational probes.

\section{Cosmological Constraints and Model Comparison}
\label{sec:results}

\subsection{Methodology and Data Combinations}

To constrain the cosmological models described in Section \ref{sec:models}, we conduct a comprehensive Bayesian parameter inference. The joint likelihood function, defined as the product of the likelihoods from each dataset ($\mathcal{L}_{\mathrm{SNIa}}$, \,$\mathcal{L}_{\mathrm{BAO}}$,\,$\mathcal{L}_{\mathrm{CMB + lensing}}$), is given by:
\begin{equation}
\mathcal{L}(\mathbf{p})=\prod_i \mathcal{L}_i. \label{eq:likelihood}
\end{equation}
The overall fit is quantified by the effective chi-squared, $\chi_{\mathrm{eff}}^2=-2 * \ln \mathcal{L}$. According to equation (\ref{eq:likelihood}), the likelihood $\mathcal{L}(\theta) \equiv P(d \mid \theta, M)$ constitutes the probability of the observed data $d$ for a specified parameter set $\theta$ within the model $M$, where it is treated strictly as a function of $\theta$.

We employ the cosmological parameter inference package \texttt{Cobaya} \citep{torrado2021cobaya, ascl:1910.019} coupled with the nested sampling algorithm \texttt{PolyChord} \citep{handley2015polychorda, handley2015polychordb} to explore the high-dimensional posterior distributions. Theoretical predictions for cosmological observables, including cosmic distances and CMB power spectra, are generated using the Einstein-Boltzmann solver \texttt{CAMB} \citep{lewis2000efficient, howlett2012cmb}, which interfaces directly with \texttt{Cobaya}.

The observational datasets utilized in this analysis comprise:
\begin{itemize}
\item \textbf{CMB+lensing}: \textit{Planck} 2018 temperature and polarization anisotropy measurements combined with CMB lensing potential power spectrum reconstructions from \textit{Planck} and ACT DR6.
\item $\boldsymbol{r_\mathrm{d}}$\textbf{(CMB)}: The comoving sound horizon at the drag epoch, $r_\mathrm{d}(\mathrm{CMB})$, is taken as a fixed value of $147.09$ Mpc. This value is derived from the TT,TE,EE+lowE+lensing constraint in the \textit{Planck} 2018 analysis (see Table 2 of \citet{2020A&A...641A...6P}).\footnote{When BAO and the full CMB power spectrum are jointly analyzed, $r_\mathrm{d}$ is determined self-consistently within the model fit. However, when BAO data are used alone, an external $r_\mathrm{d}$ value — such as $r_\mathrm{d}$(CMB) — is required to convert relative distance measurements into absolute cosmological constraints.}
\item \textbf{BAO}: A comprehensive set of 13 BAO distance measurements from DESI DR2, spanning the redshift range $0.295 \leq z_{\mathrm{eff}} \leq 2.330$.
\item \textbf{SNIa}: The Pantheon+ sample of SNIa with a cutoff of  $z> 0.01$.
\end{itemize}

We adopt uniform priors for the base cosmological parameters, consistent with established conventions in \textit{Planck} analyses \citep{2020A&A...641A...6P}: $\Omega_b h^2 \sim \mathcal{U}[0.005, 0.1]$, $\Omega_c h^2 \sim \mathcal{U}[0.001, 0.99]$, $H_0 \sim \mathcal{U}[40, 100]$, and $\tau \sim \mathcal{U}[0.01, 0.8]$. The priors for $n_s$ and $\ln(10^{10}A_s)$ are set to $\mathcal{U}[0.8, 1.2]$ and $\mathcal{U}[1.61, 3.91]$, respectively. Spatial curvature is fixed to zero ($\Omega_k=0$) throughout, corresponding to a flat universe geometry. These foundational priors remain consistent across all five cosmological models investigated.

Model-specific priors for dark energy parameters are defined as follows:
\begin{itemize}
\item $\Lambda$CDM: The dark energy EoS parameter is fixed to $w = -1$.
\item $w$CDM: A uniform prior $w \sim \mathcal{U}(-3, 0)$ is implemented.
\item $w_0w_a$CDM: Uniform priors $w_0 \sim \mathcal{U}(-3, 1)$ and $w_a \sim \mathcal{U}(-3, 2)$ are adopted.
\item $\phi$CDM: The potential steepness parameter follows $\alpha \sim \mathcal{U}(0.00001, 10)$.
\item $\xi$-index model: Uniform priors $w_\mathrm{X} \sim \mathcal{U}(-3, 0)$ and $\xi \sim \mathcal{U}(2, 10)$ are employed for the dark energy EoS and coupling parameter, respectively.
\end{itemize}

These prior ranges are deliberately expansive, encompassing or exceeding parameter spaces typically explored in contemporary literature \citep{adame2025desi, park2018observational, yan2025investigating}, thereby ensuring robust exploration of the parameter space without significant influence from prior boundaries.

\subsection{Bayesian Model Comparison}
\label{subsec:logB}

We perform a comprehensive Bayesian model comparison to assess how well five cosmological models fit current observational data. The Bayesian evidence, denoted $B \equiv P(d \mid M)$, quantifies the total probability of the observed data $d$ under model $M$, marginalized over all model parameters. Formally, it is computed by integrating the product of the likelihood $\mathcal{L}(\theta)$ and the prior distribution $\pi(\theta)$ across the entire parameter space: $B=\int \mathcal{L}(\theta) \pi(\theta) d \theta$. As a scalar measure that averages a model’s predictive capability over its parameter space, the evidence provides a principled basis for model comparison and naturally incorporates Occam’s razor—penalizing needless complexity. Following the usual scale in cosmology \citep{trotta2008bayes}, we interpret the evidence difference $\Delta \ln B$ as follows:
$(0, 1.0)$ -- inconclusive;
$(1.0, 2.5)$ -- weak evidence;
$(2.5, 5.0)$ -- moderate evidence;
$(5.0, \infty)$ -- strong evidence in favor of the model with higher evidence.

\begin{figure}[h]
    \centering
    \begin{minipage}[b]{0.45\textwidth}
        \centering
        \includegraphics[width=\textwidth]{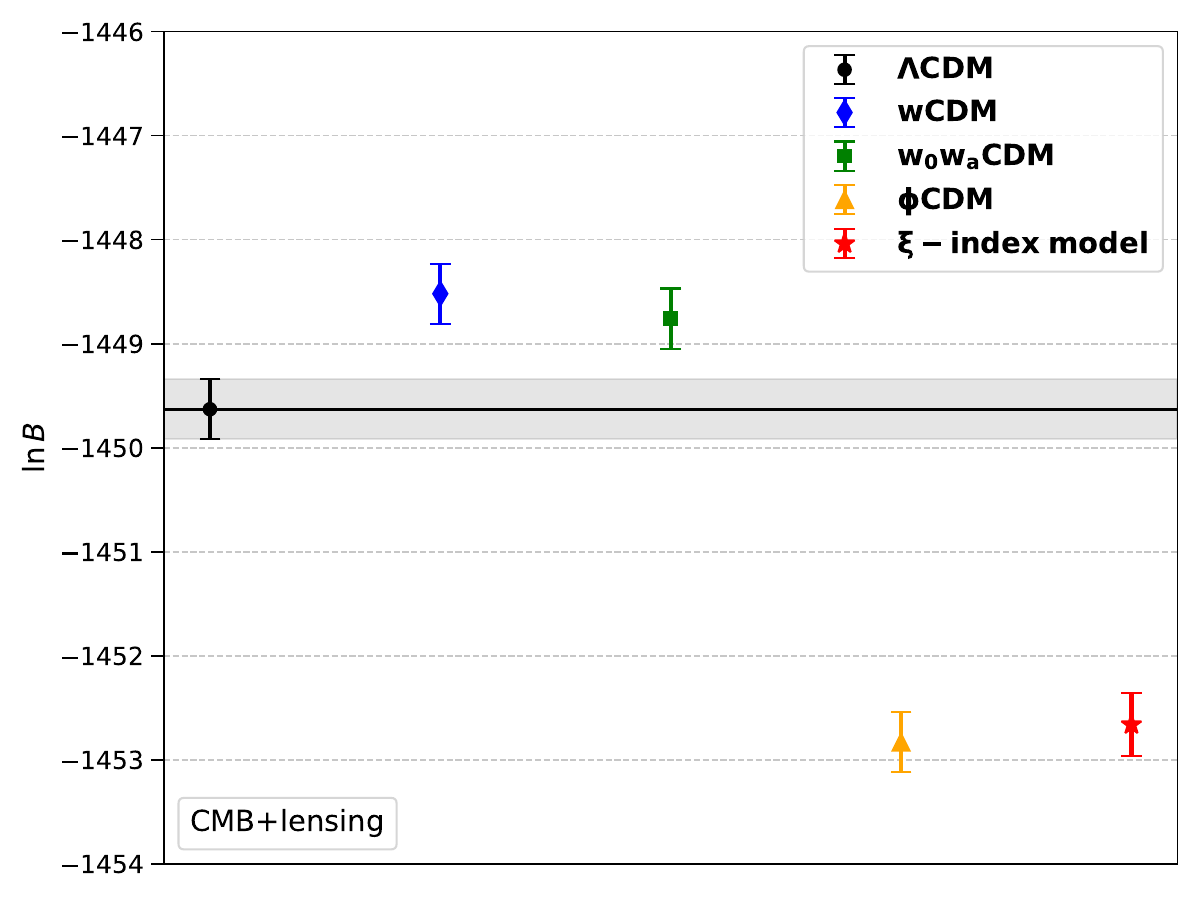}
        \small 
    \end{minipage}
    \begin{minipage}[b]{0.45\textwidth}
        \centering
        \includegraphics[width=\textwidth]{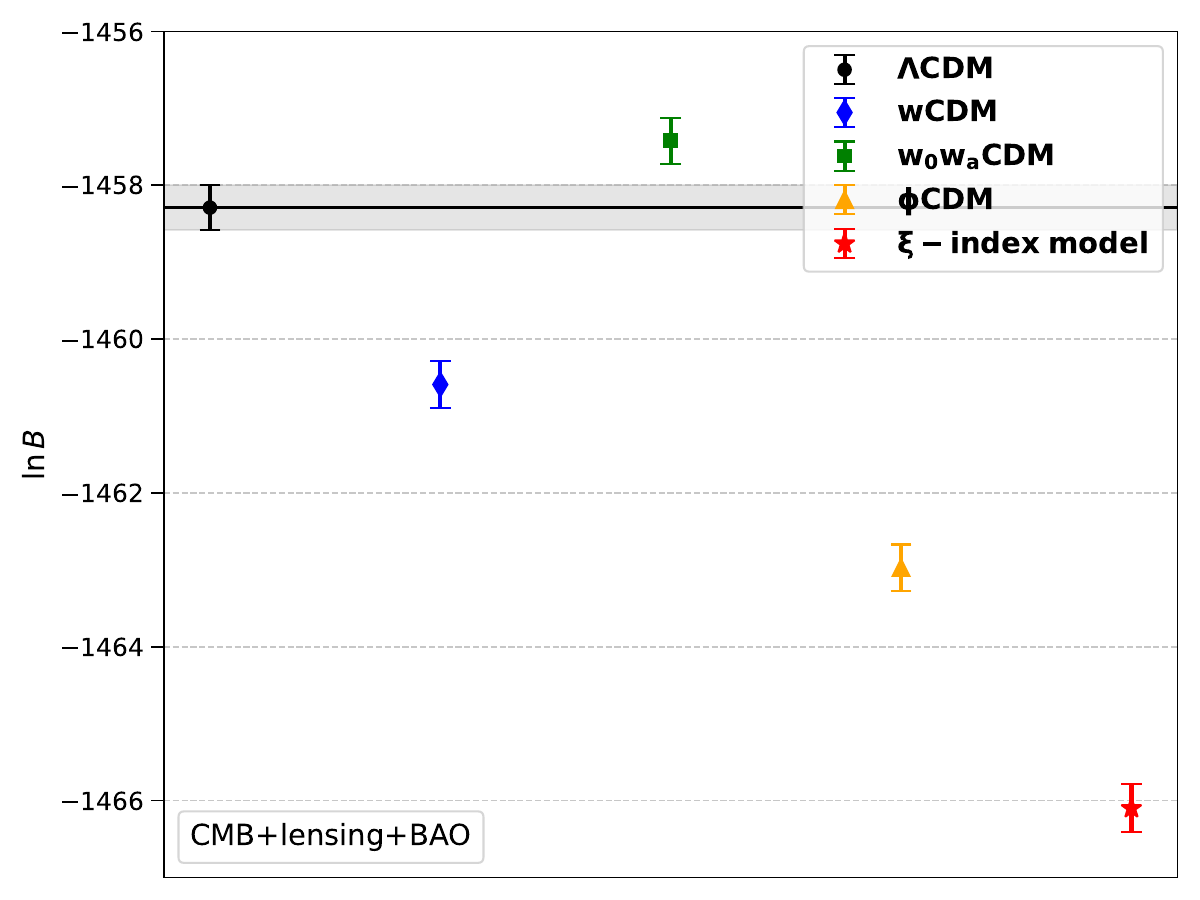}
        \small
    \end{minipage}

    \vspace{0cm}

    \begin{minipage}[b]{0.45\textwidth}
        \centering
        \includegraphics[width=\textwidth]{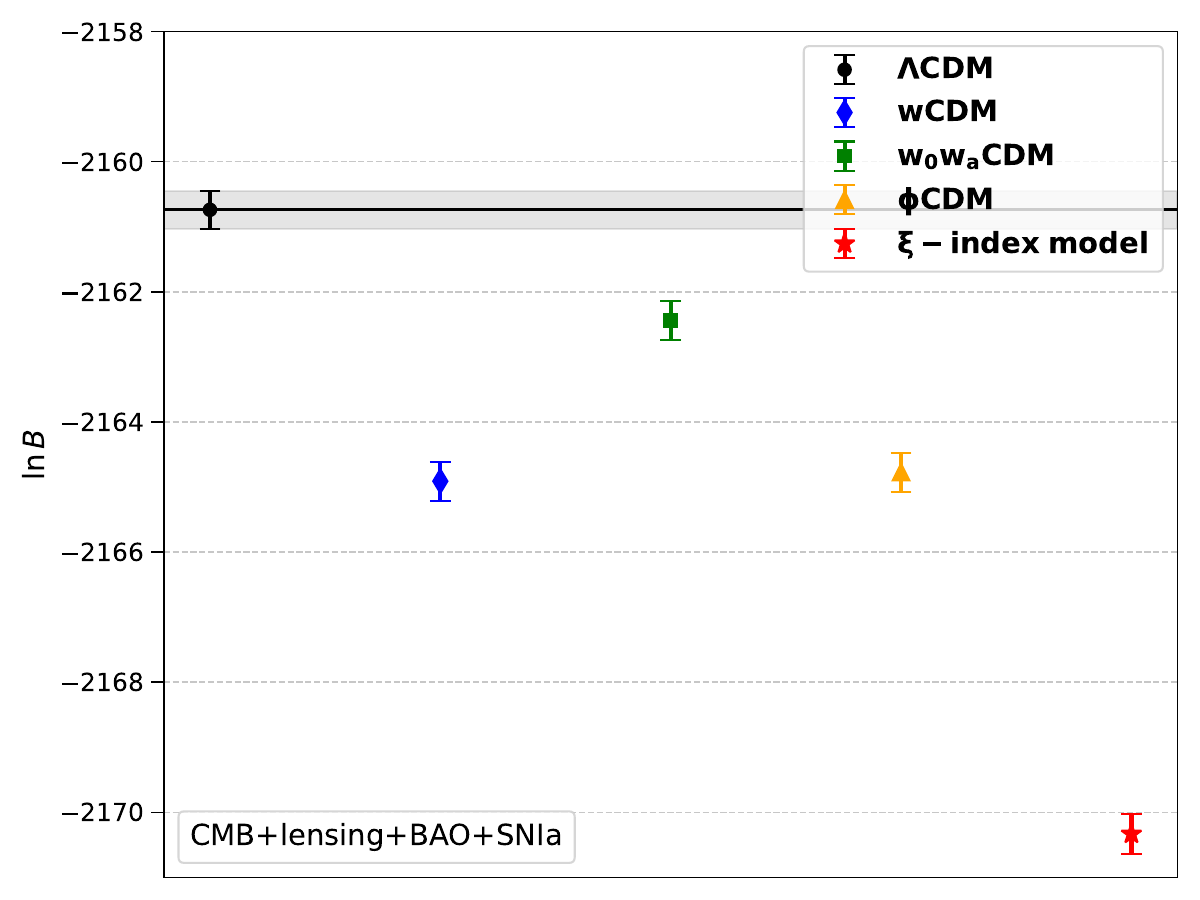}
        \small 
    \end{minipage}
    \begin{minipage}[b]{0.45\textwidth}
        \centering
        \includegraphics[width=\textwidth]{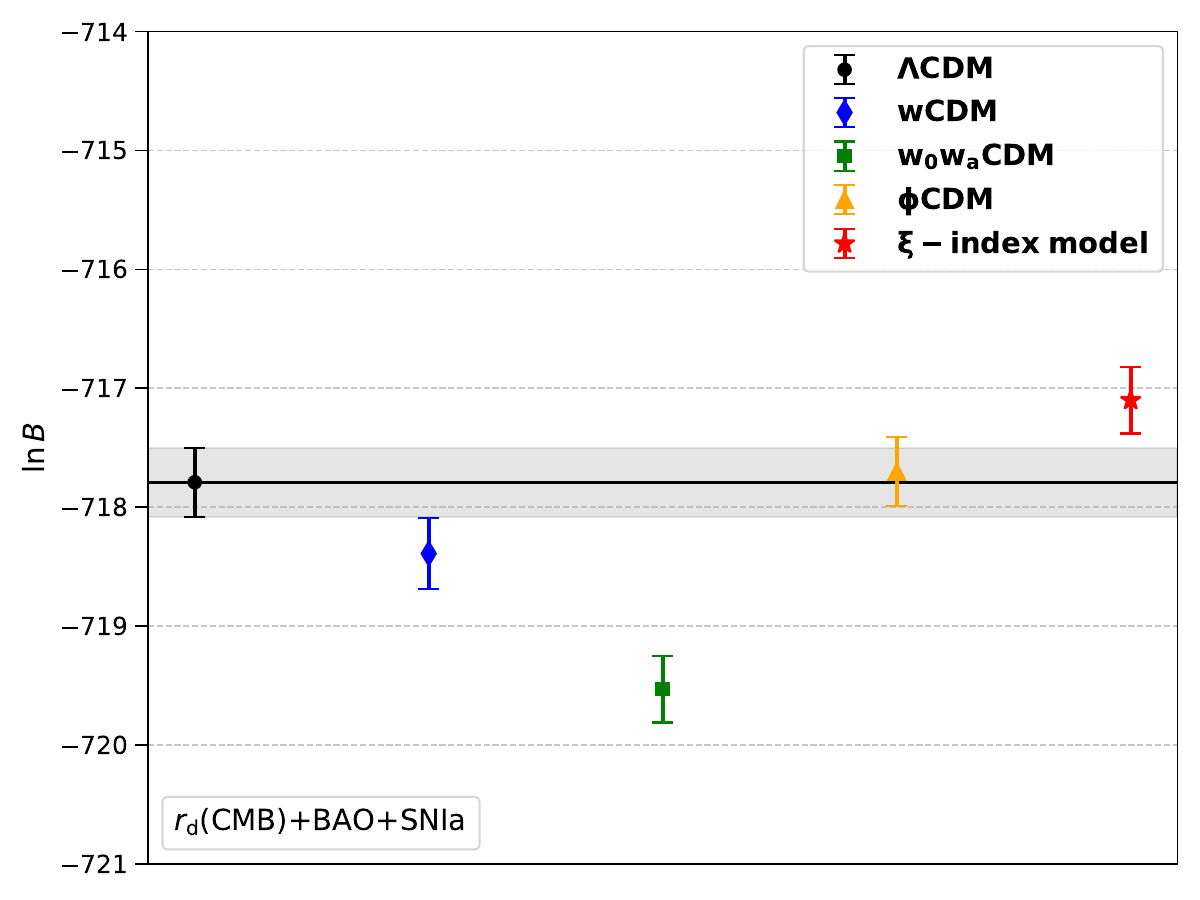}
        \small
    \end{minipage}
    \caption{Bayesian evidence comparison reveals strong dataset dependence. The $\ln B$ values for the five cosmological models are shown for four data combinations. No alternative model achieves decisive evidence ($\Delta\ln B > 2.5$) over $\Lambda$CDM (black) across all datasets. Error bars show 1$\sigma$ uncertainties.}
\label{fig:logB}
\end{figure}

Table~\ref{tab:params_summary} lists the Bayesian evidence values and parameter estimates for all models and datasets. A key result is that model preference strongly depends on the data combination used, as shown in Figure~\ref{fig:logB}. For CMB+lensing data alone, the $w$CDM model is preferred. Adding BAO data shifts support toward $w_0w_a$CDM. The full dataset (CMB+lensing+BAO+SNIa) favors the standard $\Lambda$CDM model, while $r_\mathrm{d}$\textsc{(CMB)}+BAO+SNIa most supports the interacting dark energy model ($\xi$-index).

Importantly, none of the extended models achieves $\Delta \ln B > 2.5$ over $\Lambda$CDM across all datasets, meaning no alternative model is strongly preferred. This shows that current data cannot clearly distinguish between these dark energy descriptions. The $w$CDM and $w_0w_a$CDM models perform consistently better with CMB+lensing data than $\phi$CDM and the $\xi$-index model, pointing to possible tensions between the latter models and high-redshift CMB observations.

The strong dependence of model rankings on the data highlights the need for multiple observational probes in cosmology. It also confirms that present data are not yet sufficient to clearly point to physics beyond the $\Lambda$CDM model using Bayesian evidence alone. It is important to note that this outcome does not reflect the strength of hints for dynamical dark energy—a topic that will be discussed in detail in Section \ref{subsec:w_z}. Similar findings have been reported in the works of \citet{2025JCAP...09..031C} and \citet{2025PhRvD.112d3513S}: although their respective models detected significant signals suggestive of dynamical dark energy, their calculated Bayesian evidence values exhibited substantial variations across different data combinations. Notably, when employing the same dataset as used in our study, this inconsistency persisted, with no definitive Bayesian preference for dynamical dark energy models over the $\Lambda$CDM framework.

\begin{table}
    \centering
    \setlength{\tabcolsep}{3.3pt}
    \begin{tabular}{lcccccc}
        \toprule
        \textbf{Model}/Dataset  \; \; \; \; & $\Omega_\mathrm{m0}$ & $H_0 \, [\mathrm{km \, s^{-1} \, Mpc^{-1}}]$ & \begin{tabular}{l}
            Parameters for DE
        \end{tabular} \; \; \; \; & $\chi^2_{\rm{min}}$ & & $\ln B$ \\
        \midrule
        \textbf{$\boldsymbol{\Lambda}$CDM} & & & & & &\\
        CMB+lensing & $0.316^{+0.007}_{-0.007}$ & $67.28^{+0.52}_{-0.52}$ & $w = -1$ (fixed) \hspace*{0.3cm} & $2800.0$ & & $-1449.63 \pm 0.29$ \\[7pt]
        CMB+lensing+BAO & $0.301^{+0.004}_{-0.004}$ & $68.44^{+0.30}_{-0.30}$ & $w = -1$ (fixed) \hspace*{0.3cm} & $2815.8$ & & $-1458.29 \pm 0.29$ \\[7pt]
        CMB+lensing+BAO+SNIa & $0.303^{+0.004}_{-0.004}$ & $68.31^{+0.29}_{-0.29}$ & $w = -1$ (fixed) \hspace*{0.3cm} & $4221.7$ & & $-2160.74 \pm 0.29$ \\[7pt]
        $r_\mathrm{d}$(CMB)+BAO+SNIa & $0.304^{+0.008}_{-0.008}$ & $68.69^{+0.46}_{-0.46}$ & $w = -1$ (fixed) \hspace*{0.3cm} & $1416.2$ & & $-717.79 \pm 0.29$ \\[7pt]
        \midrule
        \textbf{$\boldsymbol{w}$CDM} & & & & & &\\
        CMB+lensing & $0.195^{+0.013}_{-0.053}$ & $> 83.10$ & $w = -1.590^{+0.140}_{-0.340}$ \hspace*{0.1cm} & $2796.6$ & & $-1448.52 \pm 0.29$ \\[7pt]
        CMB+lensing+BAO & $0.292^{+0.007}_{-0.007}$ & $69.61^{+0.94}_{-0.94}$ & $w = -1.050^{+0.038}_{-0.038}$ \hspace*{0.1cm} & $2815.7$ & & $-1460.59 \pm 0.30$ \\[7pt]
        CMB+lensing+BAO+SNIa & $0.304^{+0.005}_{-0.005}$ & $68.11^{+0.59}_{-0.59}$ & $w = -0.991^{+0.023}_{-0.023}$ \hspace*{0.1cm} & $4222.5$ & & $-2164.91 \pm 0.30$ \\[7pt]
        $r_\mathrm{d}$(CMB)+BAO+SNIa & $0.298^{+0.008}_{-0.008}$ & $67.83^{+0.58}_{-0.58}$ & $w = -0.915^{+0.037}_{-0.037}$ \hspace*{0.1cm} & $1411.5$ & & $-718.39 \pm 0.30$ \\[7pt]
        \midrule
        \textbf{$\boldsymbol{w_0 w_a}$CDM} & & & & & &\\
        CMB+lensing & $0.199^{+0.010}_{-0.056}$ & $> 82.90$ & 
        \begin{tabular}{l}
            $w_0 = -1.310^{+0.420}_{-0.420}$ \\
            $w_a < -0.69$
        \end{tabular} \hspace*{0.7cm} & $2796.6$ & & $-1448.76 \pm 0.29$ \\[17pt]
        CMB+lensing+BAO & $0.350^{+0.021}_{+0.021}$ & $63.90^{+1.70}_{-2.10}$ & 
        \begin{tabular}{l}
            $w_0 = -0.430^{+0.210}_{-0.210}$ \\
            $w_a = -1.71^{+0.60}_{-0.60}$
        \end{tabular} \hspace*{0.7cm} & $2807.9$ & & $-1457.42 \pm 0.30$ \\[17pt]
        CMB+lensing+BAO+SNIa & $0.311^{+0.006}_{-0.006}$ & $67.67^{+0.58}_{-0.58}$ & 
        \begin{tabular}{l}
        $w_0 = -0.841^{+0.053}_{-0.053}$ \\
        $w_a = -0.60^{+0.22}_{-0.19}$ 
        \end{tabular} \hspace*{0.7cm} & $4214.1$ & & $-2162.44 \pm 0.30$ \\[17pt]
        $r_\mathrm{d}$(CMB)+BAO+SNIa & $0.303^{+0.018}_{-0.014}$ & $67.81^{+0.59}_{-0.59}$ & 
        \begin{tabular}{l}
            $w_0 = -0.887^{+0.055}_{-0.061}$ \\
            $w_a = -0.22^{+0.43}_{-0.37}$
        \end{tabular} \hspace*{0.7cm} & $1411.3$ & & $-719.53 \pm 0.28$ \\[14pt]
        \midrule
        \textbf{$\boldsymbol{\phi}$CDM} & & & & & &\\
        CMB+lensing & $0.329^{+0.010}_{-0.014}$ & $66.00^{+1.30}_{-0.80}$ & $\alpha < 0.136$ \hspace*{1.35cm} & $2801.6$ & & $-1452.83 \pm 0.29$ \\[7pt]
        CMB+lensing+BAO & $0.303^{+0.004}_{-0.004}$ & $68.10^{+0.43}_{-0.34}$ & $\alpha < 0.042$ \hspace*{1.35cm} & $2817.4$ & & $-1462.97 \pm 0.30$ \\[7pt]
        CMB+lensing+BAO+SNIa & $0.305^{+0.004}_{-0.004}$ & $67.91^{+0.44}_{-0.34}$ & $\alpha < 0.056$ \hspace*{1.35cm} & $4221.9$ & & $-2164.77 \pm 0.30$ \\[7pt]
        $r_\mathrm{d}$(CMB)+BAO+SNIa & $0.294^{+0.009}_{-0.009}$ & $67.79^{+0.58}_{-0.58}$ & $\alpha = 0.300^{+0.130}_{-0.160}$ \hspace*{0.5cm} & $1411.7$ & & $-717.70 \pm 0.29$ \\[7pt]
        \midrule
        \textbf{$\boldsymbol{\xi}$-index model} & & & & & &\\
        CMB+lensing & $0.191^{+0.011}_{-0.049}$ & $>84.40$ & 
        \begin{tabular}{l}
            $w_\mathrm{X} = -1.610^{+0.120}_{-0.320}$ \\
            $\xi = 4.820^{+0.960}_{-0.380}$ 
        \end{tabular} \hspace*{0.7cm} & $2797.7$ & & $-1452.66 \pm 0.30$ \\[17pt]
        CMB+lensing+BAO & $0.294^{+0.007}_{-0.007}$ & $69.53^{+0.91}_{-0.91}$ & 
        \begin{tabular}{l}
            $w_\mathrm{X} = -1.048^{+0.039}_{-0.034}$ \\
            $\xi = 3.130^{+0.100}_{-0.120}$ 
        \end{tabular} \hspace*{0.6cm} & $2815.8$ & & $-1466.1 \pm 0.31$ \\[17pt]
        CMB+lensing+BAO+SNIa & $0.305^{+0.005}_{-0.005}$ & $68.08^{+0.55}_{-0.55}$ & 
        \begin{tabular}{l}
            $w_\mathrm{X} = -0.989^{+0.022}_{-0.022}$ \\
            $\xi = 2.954^{+0.066}_{-0.066}$ 
        \end{tabular} \hspace*{0.6cm} & $4222.2$ & & $-2170.33 \pm 0.31$ \\[17pt]
        $r_\mathrm{d}$(CMB)+BAO+SNIa & $0.310^{+0.120}_{-0.170}$ & $67.82^{+0.58}_{-0.58}$ & 
        \begin{tabular}{l}
            $w_\mathrm{X} = -0.990^{+0.320}_{-0.130}$ \\
            $\xi = 2.830^{+0.460}_{-0.670}$ 
        \end{tabular} \hspace*{0.6cm} & $1411.3$ & & $-717.10 \pm 0.28$ \\[14pt] 
        \bottomrule
    \end{tabular}
    \caption{Summary of cosmological parameters, minimum $\chi^2$, and Bayesian evidence ($\ln B$). Constraints (68\% confidence level (CL)) are shown for the five models across different data combinations, including $\Omega_\mathrm{m0}$, $H_0$, and model-specific dark energy parameters.}
\label{tab:params_summary}
\end{table}

\clearpage

\subsection{Hubble Constant Constraints and the Hubble Tension}
\label{subsec:H0}

\begin{figure}[H]
    \centering
    \begin{minipage}[b]{0.45\textwidth}
        \centering
        \includegraphics[width=\textwidth]{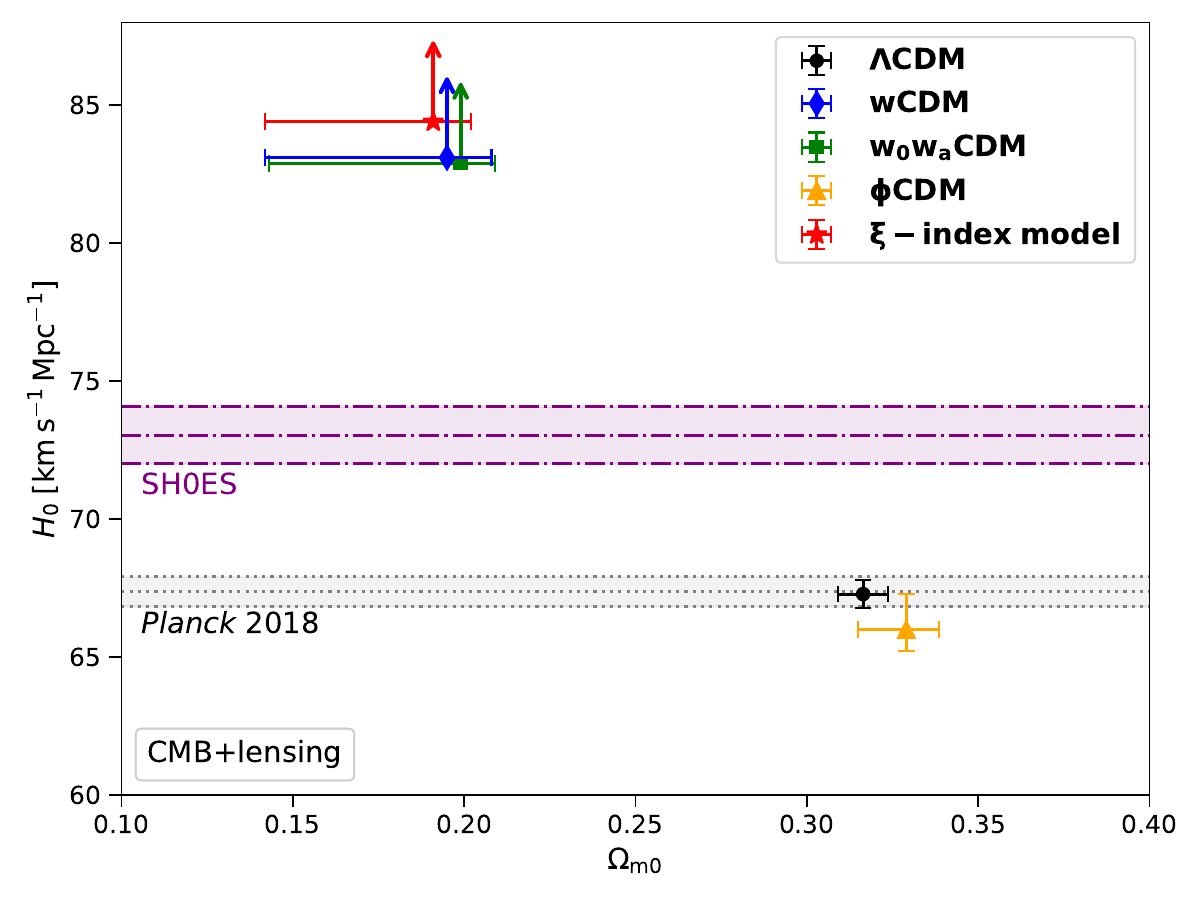}
        \small 
    \end{minipage}
    \begin{minipage}[b]{0.45\textwidth}
        \centering
        \includegraphics[width=\textwidth]{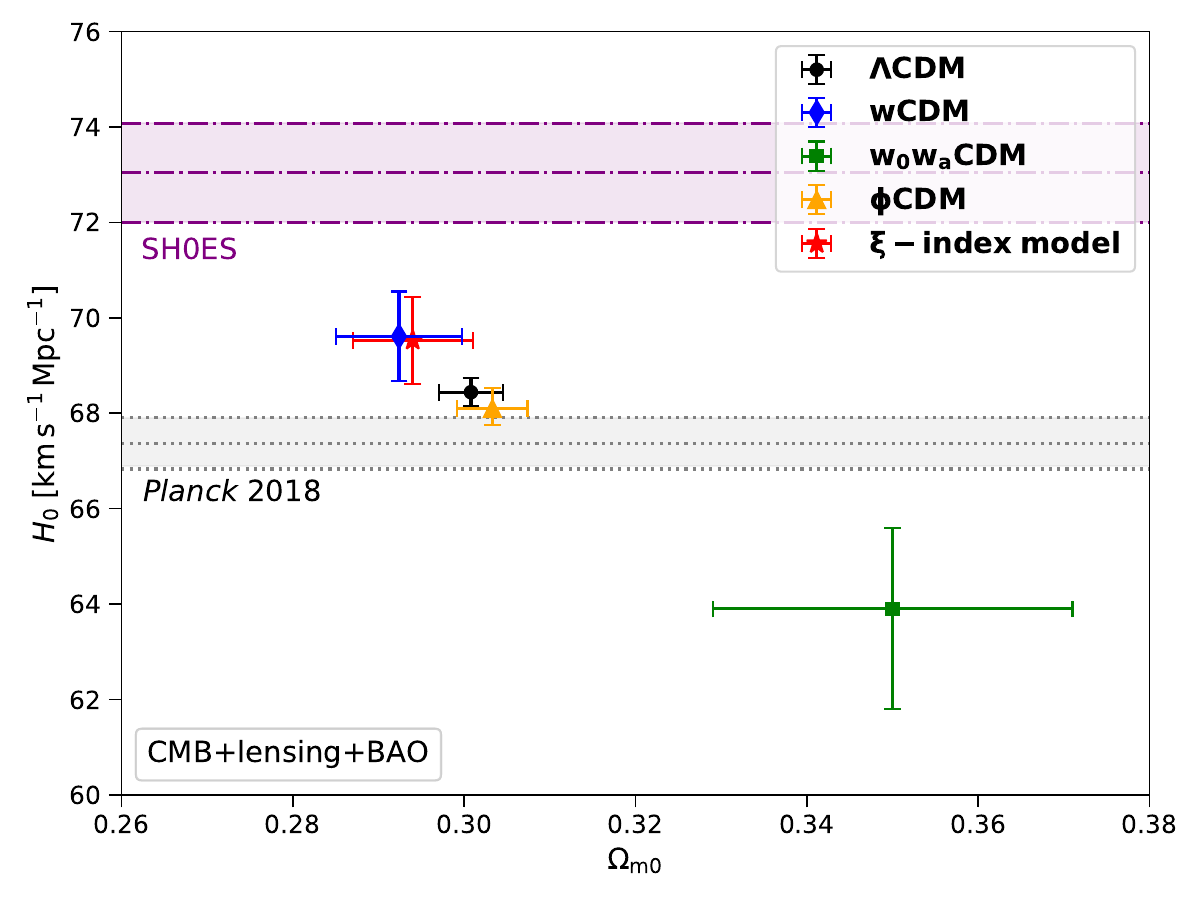}
        \small
    \end{minipage}

    \vspace{0.0cm}

    \begin{minipage}[b]{0.45\textwidth}
        \centering
        \includegraphics[width=\textwidth]{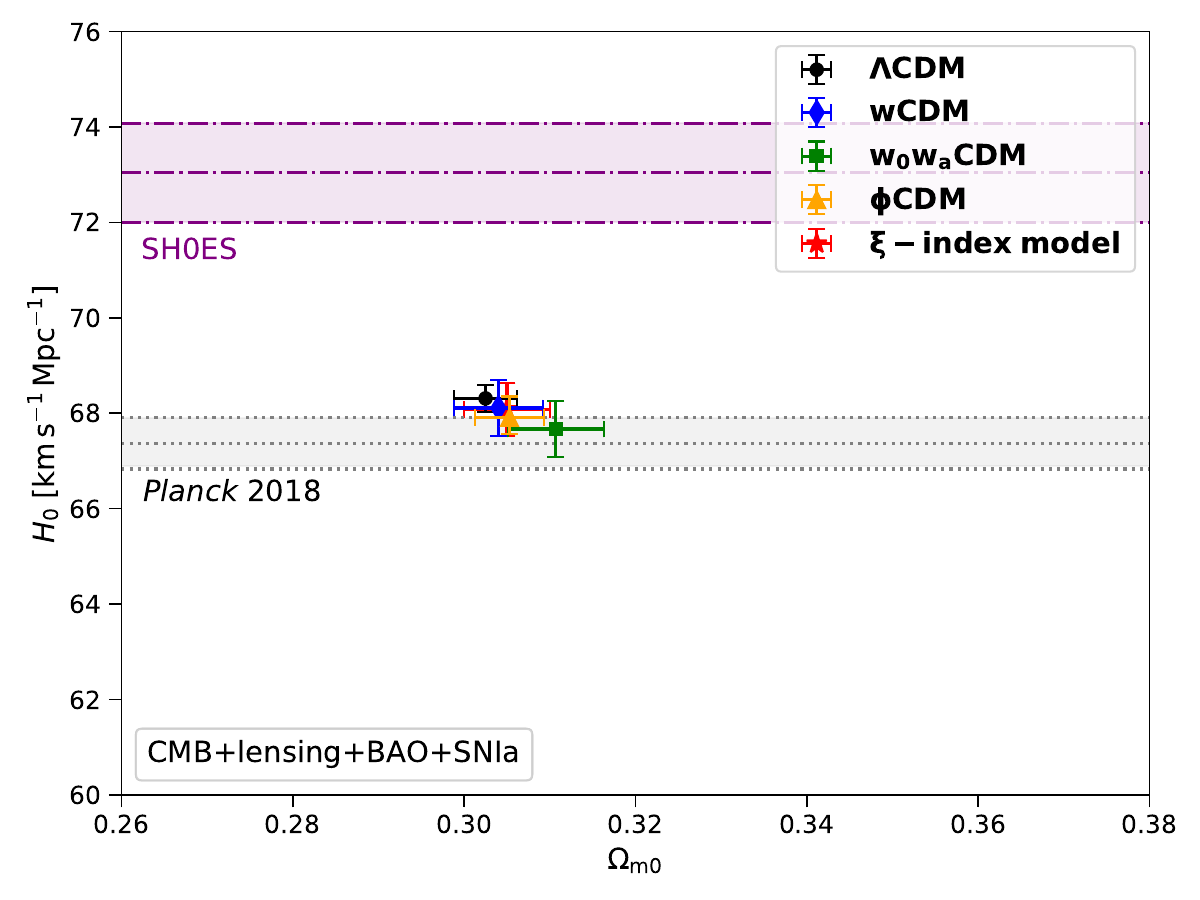}
        \small 
    \end{minipage}
    \begin{minipage}[b]{0.45\textwidth}
        \centering
        \includegraphics[width=\textwidth]{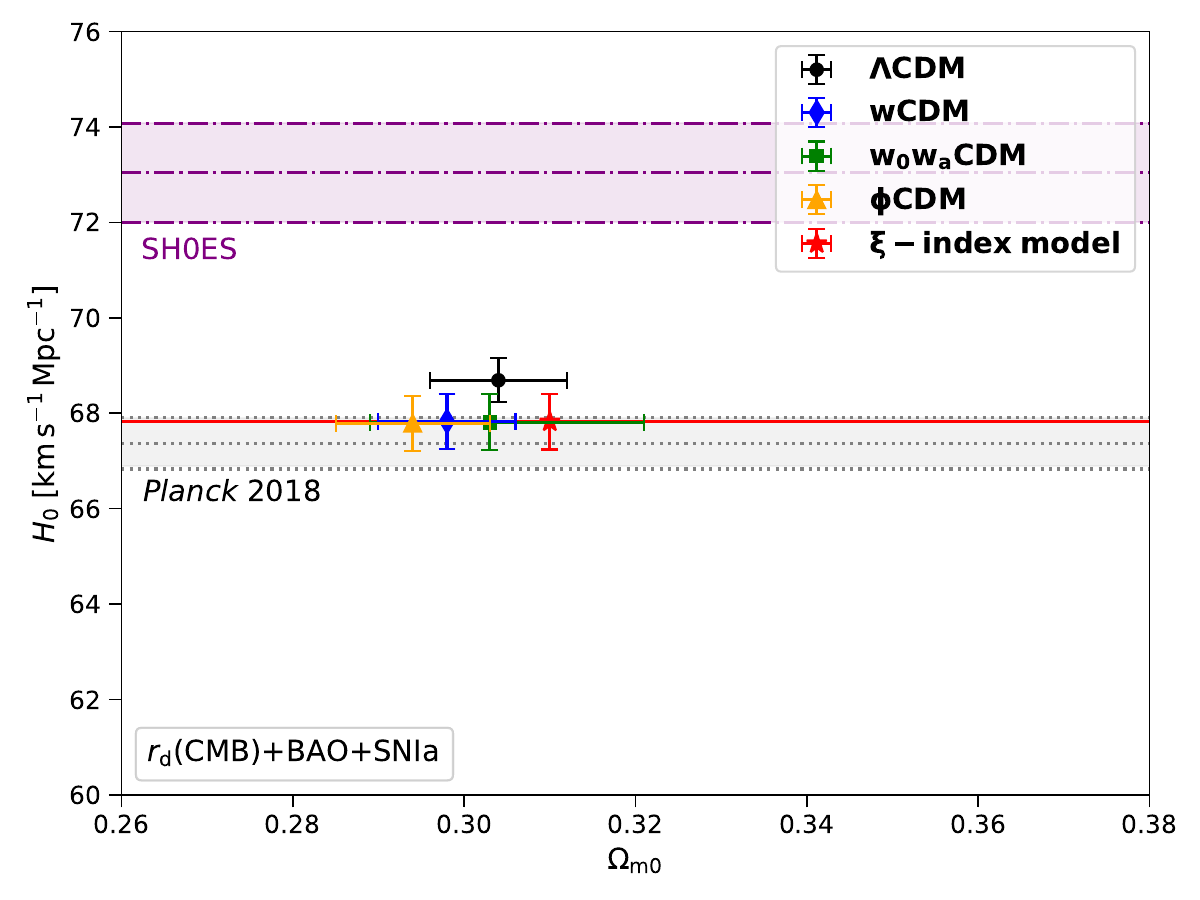}
        \small
    \end{minipage}
    \caption{Constraints on $\Omega_{\mathrm{m0}}$ and $H_0$ highlight the persistent Hubble tension. Joint 68\% CL constraints are shown for the five models (same color scheme as Fig. \ref{fig:logB}) across four data combinations. The purple dashed and gray dot-dashed lines indicate the SH0ES and \textit{Planck} 2018 measurements, respectively. All models converge to the lower $H_0$ value with constraining datasets.}
\label{fig:Om_H0}
\end{figure}

We present comprehensive constraints on $H_0$ and $\Omega_{\mathrm{m0}}$ from a multi-model analysis, shown in Figure~\ref{fig:Om_H0}, offering key insights into the Hubble tension and potential resolutions beyond the standard model.

Using CMB+lensing data alone (Figure~\ref{fig:Om_H0}, top-left), $w$CDM and $w_0w_a$CDM yield higher $H_0$ values than both the \textit{Planck} 2018 result ($67.40 \pm 0.50$ km/s/Mpc; \citealt{collaboration2020planck}) and the SH0ES measurement ($73.04 \pm 1.04$ km/s/Mpc; \citealt{riess2022comprehensive}). This trend is consistent with the theoretical prediction from \citet{2020PhRvD.102b3518V}, which establishes that a more negative EoS parameter (i.e., enhanced phantom dark energy, $w < -1$) correlates with an increased $H_{0}$ value, and aligns with the Bayesian evidence in Section~\ref{subsec:logB}, suggesting CMB+lensing data favor phantom dark energy ($w < -1$). In contrast, $\phi$CDM cannot achieve $w < -1$ and gives $H_0$ values consistent with $\Lambda$CDM.

Adding BAO data (top-right) changes $H_0$ constraints: $w$CDM and the $\xi$-index model shift toward intermediate values, reducing tension with SH0ES to $3.5\sigma$ and $3.8\sigma$, respectively. However, this improvement is not robust. Including SNIa data (bottom-left) or replacing CMB+lensing with the $r_\mathrm{d}$(CMB) prior (bottom-right) restores $H_0$ toward the \textit{Planck} value and reinstates the tension.

Crucially, across all well-constrained datasets, none of the alternative models produce $H_0$ values matching the SH0ES measurement. This indicates that these model extensions, despite added flexibility, do not resolve the Hubble tension with current data.

The consistent $H_0$ estimates across models under strong constraints highlight the tension's severity. This suggests the issue likely originates from either unmodeled systematics or new physics beyond the tested dark energy scenarios.

\subsection{Evolution of the Dark Energy Equation of State}
\label{subsec:w_z}
\begin{figure}[H]
\centering
\begin{minipage}[b]{0.45\textwidth}
\centering
\includegraphics[width=\textwidth]{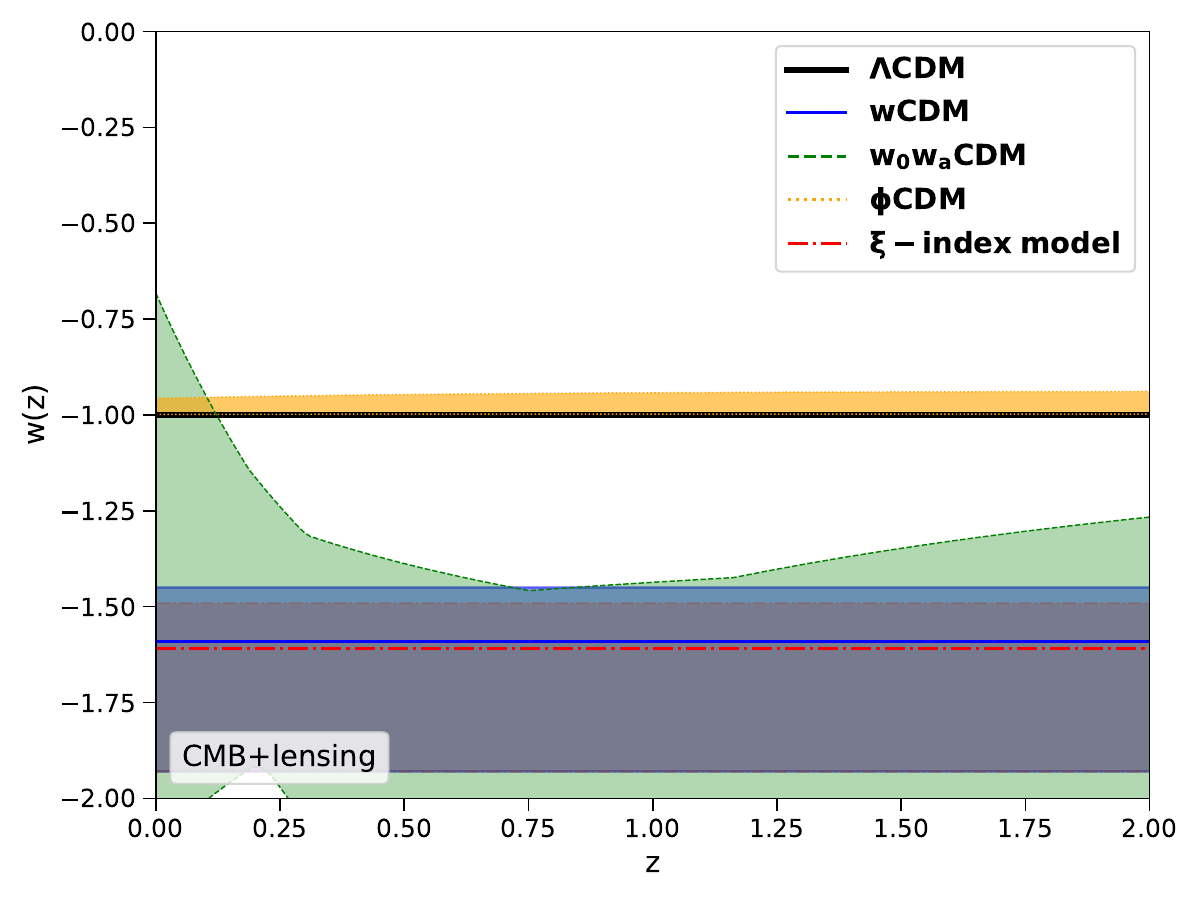}
\small
\end{minipage}
\begin{minipage}[b]{0.45\textwidth}
\centering
\includegraphics[width=\textwidth]{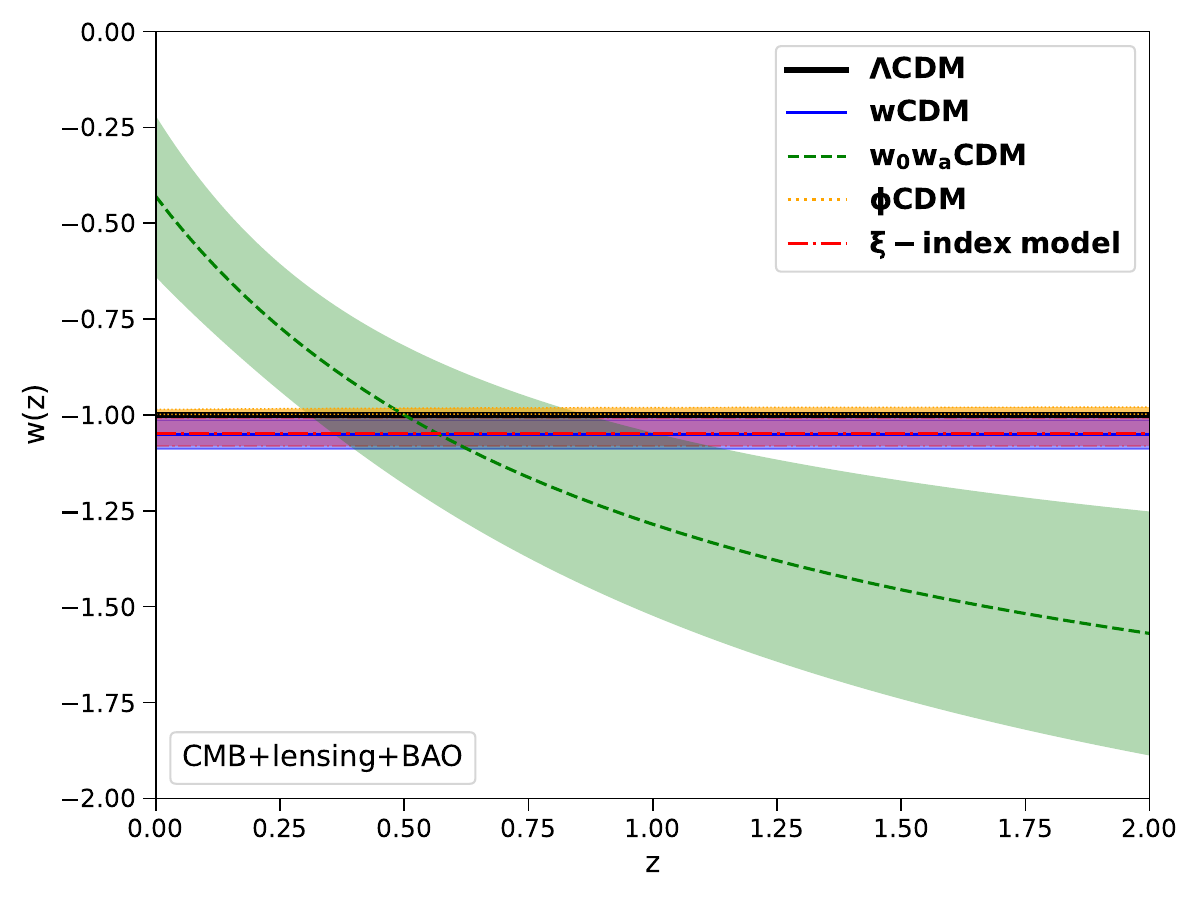}
\small
\end{minipage}
\vspace{0.0cm}

\begin{minipage}[b]{0.45\textwidth}
    \centering
    \includegraphics[width=\textwidth]{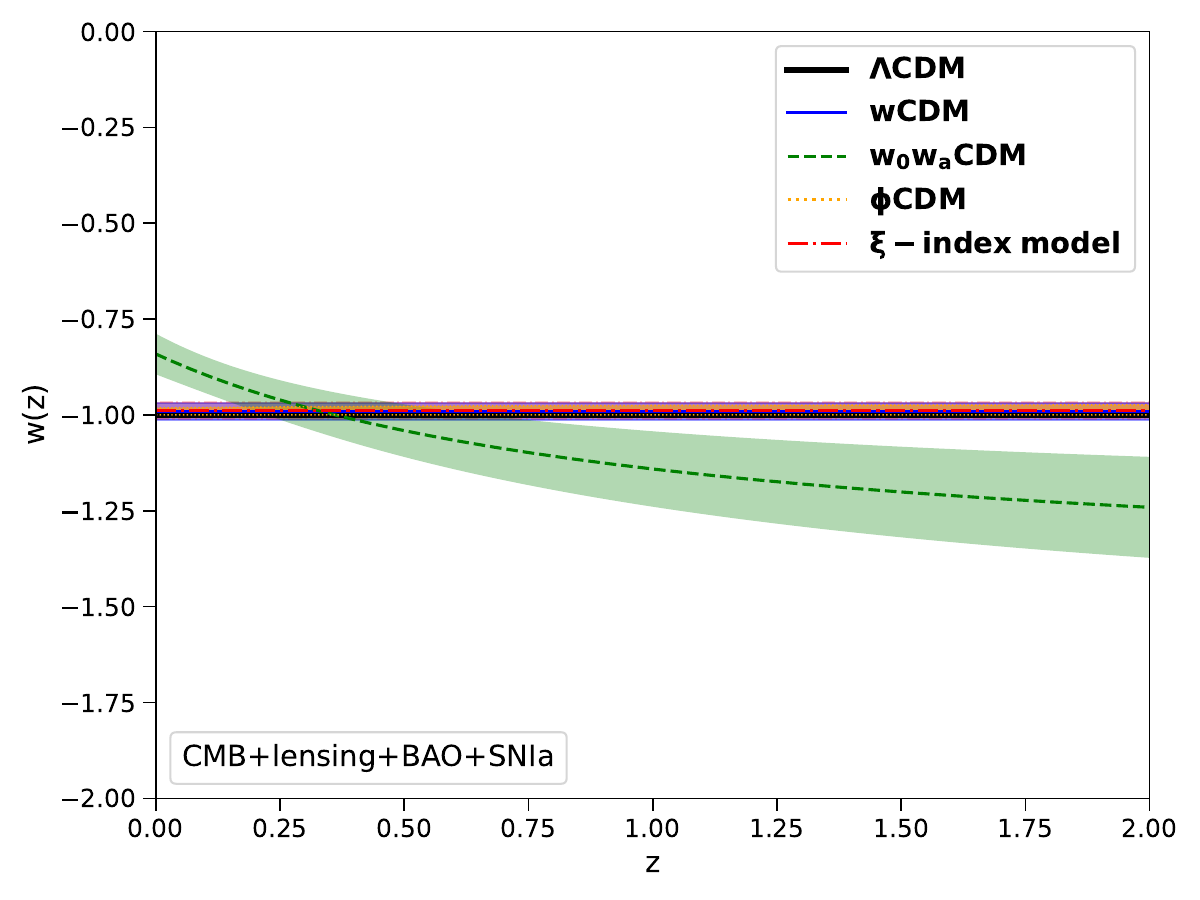}
    \small 
\end{minipage}
\begin{minipage}[b]{0.45\textwidth}
    \centering
    \includegraphics[width=\textwidth]{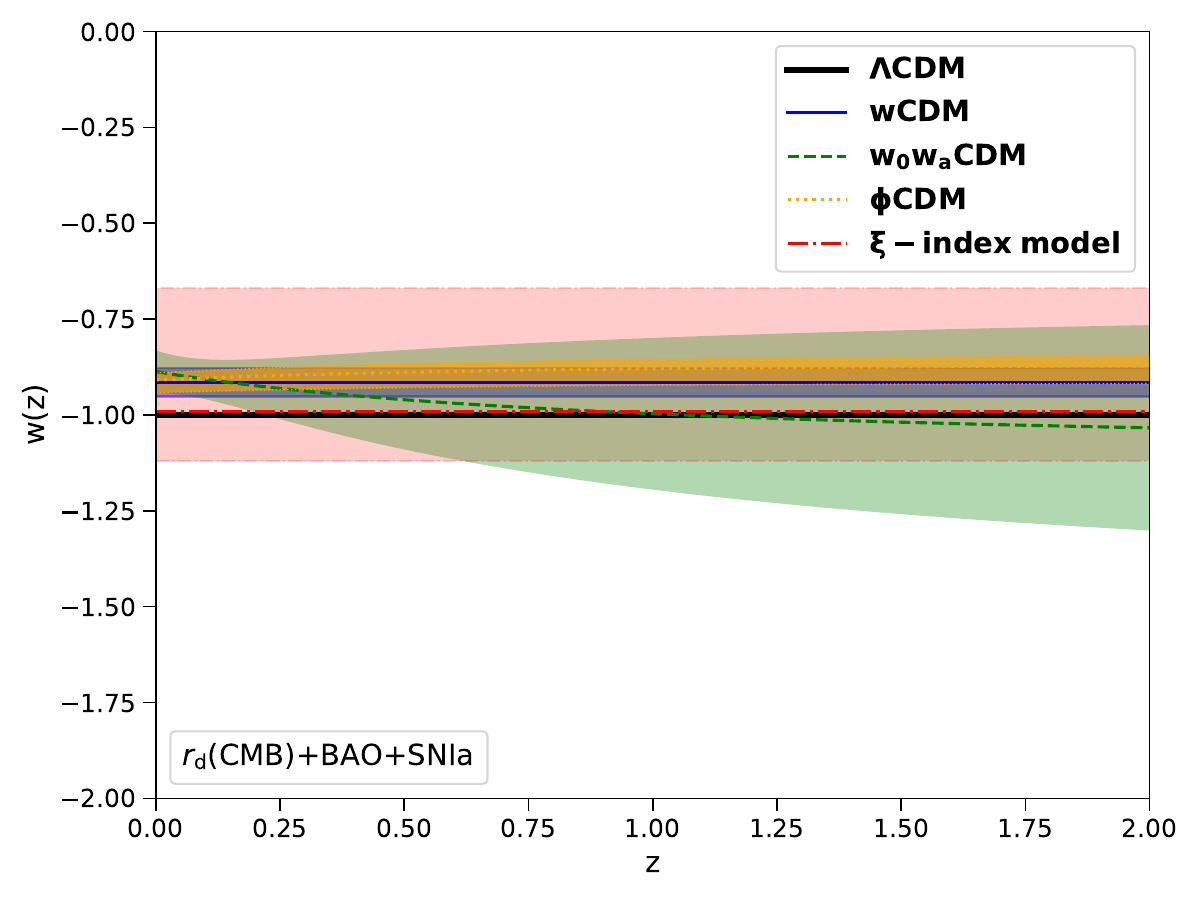}
    \small
\end{minipage}
\caption{Redshift evolution of $w(z)$ provides evidence for dynamical dark energy. The 1$\sigma$ constraints on $w(z)$ are shown for the five models (same style and color as previous figures) across four data combinations. A clear evolutionary trend is observed: high-$z$ CMB data prefer phantom behavior ($w < -1$), while low-$z$ BAO and SNIa data favor quintessence ($w > -1$).}
\label{fig:w_z}
\end{figure}

The redshift evolution of the dark energy EoS parameter, $w(z)$, provides crucial insights into cosmic acceleration. Figure~\ref{fig:w_z} shows the constraints for each model and dataset, revealing strong evidence for time-varying dark energy.

With CMB+lensing data alone (Fig.~\ref{fig:w_z}, top-left), both $w$CDM and $w_0 w_a$CDM favor phantom behavior ($w < -1$) across all redshifts. This aligns with the Bayesian evidence in Section~\ref{subsec:logB}, indicating early-universe data prefer a phantom phase. In contrast, the $\phi$CDM model yields $w(z)$ asymptotically approaching $-1$, consistent with its prohibition of phantom crossing.

Adding BAO data (top-right) alters the inferred evolution significantly. Constant $w$ in $w$CDM shifts toward $-1$, while $w_0 w_a$CDM shows a distinct transition from phantom ($w < -1$) at high redshifts to quintessence ($w > -1$) at low redshifts. This crossing of the phantom divide places $w_0 w_a$CDM within the Quintom paradigm\footnote{The Quintom model bridges Quintessence ($w > -1$) and Phantom ($w < -1$), allowing $w$ to cross $-1$.}, supporting recent independent findings.

Adding SNIa data (bottom-left) strengthens this trend for $w_0 w_a$CDM. It also pulls the mean $w(z)$ for $w$CDM, $\phi$CDM, and $\xi$-index toward quintessence at low redshifts, highlighting the role of late-time distance measurements and a persistent tension between early- and late-universe observables.

Replacing the full CMB+lensing likelihood with the $r_\mathrm{d}$(CMB) prior (bottom-right) yields less distinct features, with all non-$\Lambda$CDM models converging toward quintessence at low redshifts. Notably, $w$CDM, $w_0 w_a$CDM, and $\phi$CDM exclude $w = -1$ at present (68\% CL), providing statistical evidence against a cosmological constant.

Collectively, high-redshift CMB data favor a phantom phase, while low-redshift BAO and SNIa measurements drive constraints toward quintessence. This tension suggests either complex dark energy dynamics or residual systematics. The evidence for temporal evolution across multiple models strongly indicates dynamical dark energy.

Our parameter estimates for the $w$CDM, $w_0w_a$CDM, and $\phi$CDM models---derived from the CMB+lensing+BAO+SNIa combination---are quantitatively consistent (within 68\% CL) with recent analyses employing similar data compilations \citep{2024PhRvD.110b3506P, 2025IJMPD..3450058P, 2025arXiv250925812P}. Those studies used a ``P18+lensing+non-CMB'' dataset, comprising \textit{Planck} 2018 CMB data, \textit{Planck} lensing, and an earlier set of BAO, SNIa, and growth factor measurements, but did not include the ACT DR6 lensing or DESI DR2 BAO data incorporated here. The narrower uncertainties in our constraints demonstrate the enhanced precision gained from these newer datasets. Critically, the qualitative agreement between our results and earlier work reinforces the robustness of the evidence for dynamical dark energy evolution, indicating that this finding is not driven by any single dataset. The coherent trend for a redshift-dependent $w(z)$ across multiple models, driven by the tension between early- and late-universe probes, provides intriguing \textit{hints} for dynamical dark energy, although Bayesian model comparison does not yet favor these extensions decisively over the $\Lambda$CDM paradigm.

\subsection{Constraints on Dark Energy--Matter Interactions}
\label{subsec:xi_constraints}

\begin{figure*}[!h]
\centering
\begin{minipage}[b]{0.48\textwidth}
\centering
\includegraphics[width=\textwidth]{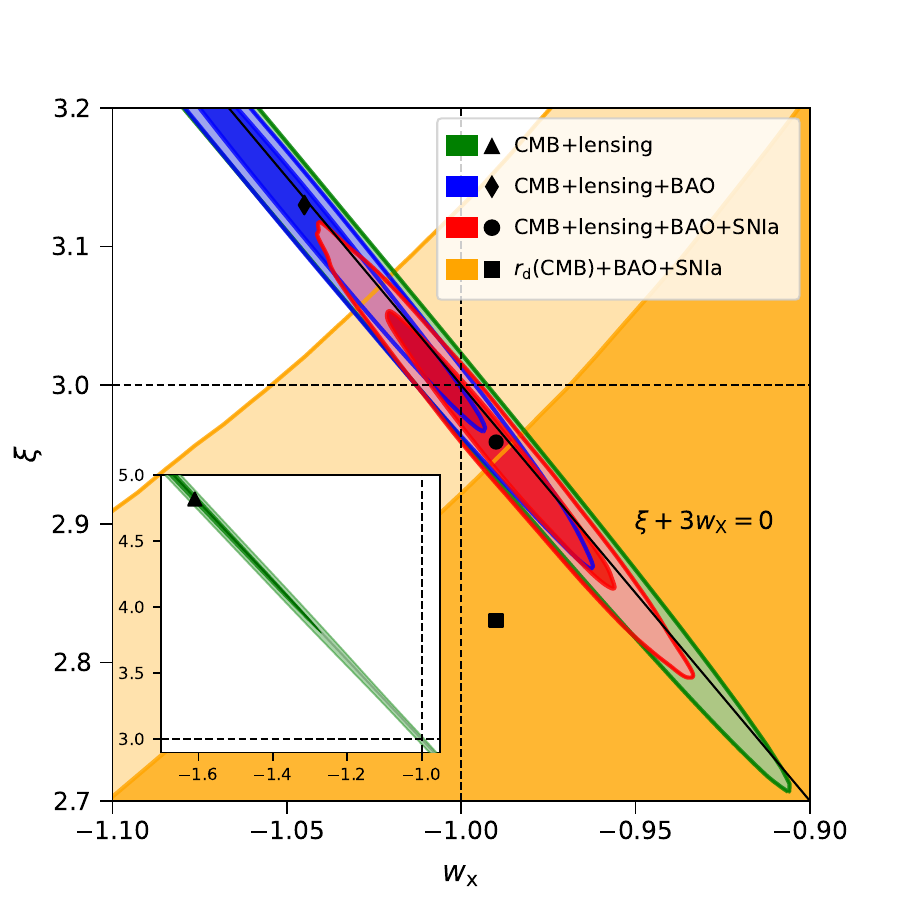}
\end{minipage}
\hfill
\begin{minipage}[b]{0.48\textwidth}
\centering
\includegraphics[width=\textwidth]{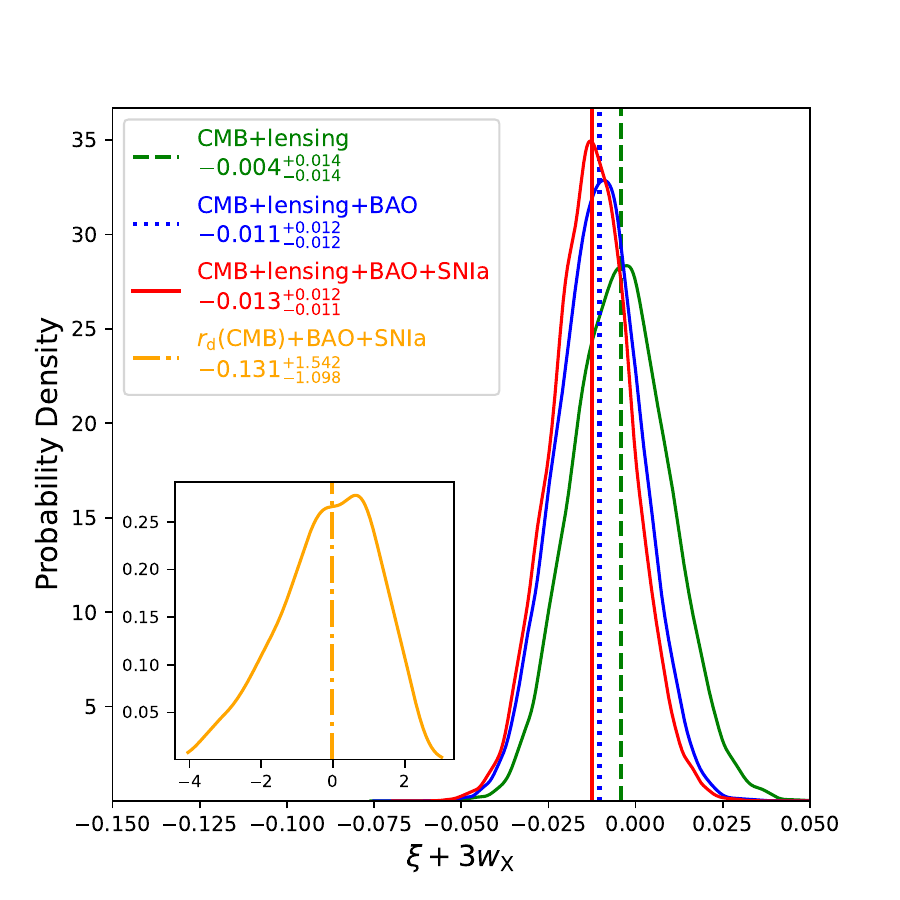}
\end{minipage}
\caption{Constraints on interacting dark energy from the $\xi$-index model. (Left) Joint 68\% and 95\% CL constraints in the $w_\mathrm{X}$-$\xi$ plane. The black solid and dashed lines mark the non-interacting and $\Lambda$CDM limits, respectively. (Right) Posterior distributions of the interaction strength $\xi + 3w_\mathrm{X}$. A negative value indicates energy transfer from dark energy to matter. The full dataset (CMB+lensing+BAO+SNIa) favors a negative coupling at 68\% CL (gray band).}
\label{fig:xi_index_model}
\end{figure*}

We analyze the interaction parameters in the $\xi$-index model, focusing on dark energy--matter coupling. The left panel of Figure~\ref{fig:xi_index_model} shows joint constraints on $w_\mathrm{X}$ and $\xi$, revealing systematic deviations from standard cosmology. Three of four dataset combinations---CMB+lensing, CMB+lensing+BAO, and $r_\mathrm{d}$(CMB)+BAO+SNIa---exclude the $\Lambda$CDM limit ($w_\mathrm{X} = -1$, $\xi = 3$) at 68\% CL. When including low-redshift probes (BAO and SNIa), $w_\mathrm{X}$ shifts toward quintessence ($w_\mathrm{X} > -1$), matching evolutionary trends in Section~\ref{subsec:w_z} and highlighting the need for multi-epoch observations.

Constraints depend strongly on the dataset. When considering CMB+lensing data alone, the non-interacting scenario ($\xi + 3w_\mathrm{X} = 0$) is not excluded; indeed, the data remain consistent with this scenario while also allowing for phantom behavior ($w_\mathrm{X} < -1$). The combined parameter $\xi + 3w_\mathrm{X}$, controlling the energy-transfer direction, clarifies the interaction. As shown in the right panel, CMB+lensing data give $\xi + 3w_\mathrm{X}$ consistent with zero, but adding low-redshift measurements shifts it to negative values. The full CMB+lensing+BAO+SNIa combination yields $\xi + 3w_\mathrm{X} < 0$ at 68\% CL, indicating energy flows from dark energy to matter in the late universe.

The $r_\mathrm{d}$(CMB)+BAO+SNIa dataset also deviates from the non-interacting case, though its weaker constraints on $\xi + 3w_\mathrm{X}$ prevent firm conclusions. Our finding of negative coupling ($\xi + 3w_\mathrm{X} < 0$) agrees with recent studies using gravitational-wave sirens \citep{zheng2022investigating} and gamma-ray bursts \citep{nong2024testing}, as well as with the results from \citet{2024ApJ...976....1L}. Using the combined dataset of CMB+DESI DR1+DESY5, they showed that all three dark energy-matter interaction models outperforming $\Lambda$CDM ($\Delta\mathrm{AIC}<0$) consistently exhibit significant energy transfer from dark energy to matter. By contrast, our conclusion differs from earlier work favoring positive coupling \citep{guo2007probing, chen2010using, cao2011testing, yan2025investigating}. This contrast highlights how constraints depend on both the model and data, underscoring the need for multi-messenger approaches.

Together, these results provide tentative but significant evidence for dark energy--matter interactions at low redshifts, with energy likely flowing from dark energy to matter. The dataset dependence of the constraints shows that future high-precision measurements across cosmic epochs are essential to properly test interacting dark energy models and understand dark-sector couplings.

\section{Summary and Conclusions} \label{sec:conclusion}
We have performed a comprehensive Bayesian analysis of five cosmological models—$\Lambda$CDM, $w$CDM, $w_0w_a$CDM, $\phi$CDM, and the interacting dark energy scenario (i.e., $\xi$-index model)—using the latest data from DESI DR2 BAO, Pantheon+ SNIa, and \textit{Planck} 2018 \& ACT DR6 CMB+lensing. Our systematic comparison yields the following principal conclusions:

\begin{itemize}

\item \textbf{Model preference is highly dataset-dependent.} Bayesian evidence comparisons show that no alternative model achieves decisive evidence ($\Delta\ln B > 2.5$) over $\Lambda$CDM model across all data combinations. The preferred model shifts with the dataset: $w$CDM model is favored by CMB+lensing alone, $w_0w_a$CDM by CMB+lensing+BAO, $\Lambda$CDM by the full dataset (including SNIa), and the $\xi$-index model when using $r_\mathrm{d}$(CMB)+BAO+SNIa. This underscores that current data are not yet sufficient to unambiguously distinguish between these competing descriptions of dark energy.

\item \textbf{The Hubble tension persists across models.} The inferred Hubble constant $H_0$ is remarkably consistent across all models when constrained by datasets that provide tight parameter bounds, invariably converging to the lower value from early-universe (\textit{Planck}) measurements. While some extended models ($w$CDM, $\xi$-index) can slightly raise $H_0$ with specific data combinations, this effect is not robust. None of the models successfully reconcile the early- and late-universe $H_0$ measurements, pointing to either more fundamental new physics or unresolved systematics.

\item \textbf{Strong evidence for dynamical dark energy emerges.} We find a clear redshift evolution in the dark energy EoS parameter. High-redshift CMB data prefer a phantom state ($w < -1$), while low-redshift BAO and SNIa data pull the constraints toward quintessence ($w > -1$). This trend is most pronounced in the $w_0w_a$CDM model, which exhibits a "Quintom"-like crossing of the phantom divide. This robust evolutionary signature strongly suggests that dark energy is dynamic and fundamentally different from a cosmological constant.

\item \textbf{Tentative evidence for late-universe interactions exists.} In the $\xi$-index model, the combination of CMB+lensing+BAO+SNIa data favors a negative coupling ($\xi + 3w_\mathrm{X} < 0$) at 68\% CL, indicating a net energy flow from dark energy to matter at late times. Although this evidence is not yet decisive, it aligns with emerging hints of dark energy--matter interactions and merits further investigation.
\end{itemize}

To summarize, our results show clear evidence for the dynamic evolution of dark energy over cosmic history. The persistent Hubble tension, unresolved by existing theoretical extensions, points to a deeper challenge in cosmology. Moving forward, it is essential to focus on next-generation multi-probe surveys, develop more sophisticated theoretical frameworks, and rigorously address systematic uncertainties to ultimately explain the mechanism of cosmic acceleration.

\begin{acknowledgments}
We thank Prof. Bharat Ratra for his insightful suggestion to include a comparative discussion of our parameter constraints with earlier related work, which strengthened the interpretation of our results in Section 4.4. This work has been supported by the National Natural Science Foundation of China (Nos.\ 12588202 and 12473002), the National Key Research and Development Program of China (Nos.\ 2022YFA1602903 and 2023YFB3002501), and the China Manned Space Program with grant no.\ CMS-CSST-2025-A03. 
\end{acknowledgments}

\vspace{5mm}

\software{\texttt{Cobaya} \citep{ascl:1910.019, torrado2021cobaya}, \texttt {CAMB} \citep{lewis2000efficient, howlett2012cmb}, \texttt {PolyChord} \citep{handley2015polychorda, handley2015polychordb}}

\bibliography{sample701}{}
\bibliographystyle{aasjournal}

\end{document}